\documentclass{wiley-article}

\usepackage[super,sort&compress]{natbib}
\setcitestyle{super,sort&compress,comma}

\usepackage[hidelinks]{hyperref}

\usepackage{siunitx}

\papertype{Research Article}

\title{
  Multi-Signal Safety Surveillance with Bayesian Latent Factor Modeling and Bias Correction
}

\author[1]{Ziyang Pan}
\author[1]{Fan Bu}

\affil[1]{Department of Biostatistics, University of Michigan, Ann Arbor, Michigan, USA}

\corraddress{Fan Bu, Department of Biostatistics, University of Michigan, Ann Arbor, Michigan, USA}
\corremail{fbu@umich.edu}

\runningauthor{Pan et al.}

\begin{document}

\begin{frontmatter}
\maketitle

\begin{abstract}

Safety surveillance increasingly involves repeated monitoring of many exposure-outcome signals in observational healthcare data, where sparse information, dependence across related signals, and systematic error can complicate inference. Existing frameworks typically focus on either correcting residual bias using negative controls or borrowing information across exposure-outcome pairs, but not both. We propose a multi-signal Bayesian sequential surveillance framework that integrates empirical bias correction with low-rank latent factor modeling. At each analysis time, a hierarchical Bayesian model learns exposure-specific bias distributions from negative control outcomes assumed to have null latent effects. Conditional on these distributions, low-rank latent factors are estimated across exposures and outcomes of interest to share information across correlated signals. As new data accrue, posterior inference is updated sequentially, yielding bias-corrected posterior summaries of effect sizes across multiple monitored signals. We illustrate the method in a postmarket vaccine safety surveillance study using a large US insurance claims database.

\keywords{bayesian sequential surveillance; empirical bias correction; latent factor model; negative control outcomes; postmarket safety surveillance}
\end{abstract}
\end{frontmatter}

\section{Introduction}

Postmarket safety surveillance is essential because rare or delayed adverse events may not be detectable before broad population use, given limited sample size, follow-up durations, and selected study populations \citep{verstraeten2003vaccine, lieu2007real, kulldorff2011maximized, madigan2014systematic}. Large linked healthcare databases, including administrative claims and electronic health records, provide an important resource for active surveillance because they contain longitudinal, time-stamped information on medical product exposures, diagnoses, and healthcare encounters across large populations \citep{verstraeten2003vaccine, madigan2014systematic, simpson2013multiple}. In many operational surveillance settings, data are analyzed repeatedly as they accrue, with the goal of detecting elevated adverse-event risk as early as possible while limiting false alarms \citep{lieu2007real, greene2011near, kulldorff2011maximized}. Sequential approaches such as the maximized sequential probability ratio test have therefore become widely used in safety monitoring, and Bayesian sequential approaches have also been proposed to provide posterior-probability-based evidence summaries and more flexible updating as data accumulate \citep{kulldorff2011maximized, li2020bayesian, bu2024bayesian}.

Despite these opportunities, safety surveillance using observational healthcare data is vulnerable to systematic error. Residual confounding, outcome misclassification, incomplete capture of exposures or outcomes, temporal changes in healthcare utilization, and data accrual delays can bias exposure-outcome effect estimates and compromise uncertainty quantification \citep{madigan2014systematic, newcomer2018bias, schuemie2014interpreting, bu2024bayesian}. Negative control outcomes and empirical calibration provide one strategy for characterizing and adjusting for such residual bias. In this framework, outcomes believed to have no causal association with the exposure are analyzed to learn the empirical distribution of residual systematic error, which can then be incorporated into inference for outcomes of interest \citep{schuemie2014interpreting, schuemie2016robust, tchetgen2014control, schuemie2018improving}. Recent Bayesian safety surveillance methods extend this idea to sequential monitoring by using negative control outcomes to inform an empirical bias distribution and updating posterior probabilities for increased adverse-event risk over time \citep{bu2024bayesian}.

A second challenge is that modern safety surveillance is inherently multi-signal. Rather than evaluating a single exposure-outcome pair in isolation, investigators often monitor multiple exposures and multiple adverse events simultaneously. Standard pairwise analyses can be unstable when individual signals are sparse and may fail to exploit shared structure across related exposures or outcomes. Self-controlled case series methods have provided a useful modeling framework for safety surveillance with longitudinal observational healthcare data, and have been extended to large-scale settings involving multiple medical product exposures and adverse-event outcomes \citep{whitaker2006tutorial, simpson2013multiple, suchard2013empirical}. These developments highlight the need for regularization, calibration, and scalable inference when many exposure-outcome associations are evaluated in large observational databases \citep{simpson2013multiple, suchard2013empirical, madigan2014systematic}.

Latent factor models offer an approach for structured borrowing across many related signals. By representing a high-dimensional association matrix through a lower-dimensional set of latent factors, these models can reduce the number of free parameters, stabilize estimation, and improve interpretability by capturing shared patterns among exposures and outcomes \citep{carvalho2008high, ghosh2009default}. The factorized self-controlled case series method introduced a hierarchical Bayesian framework for estimating multiple exposure-outcome effects simultaneously, using latent factors to borrow information across both dimensions of the exposure-outcome effect matrix \citep{moghaddass2016factorized}. However, latent factor surveillance models and empirical bias-correction methods have largely developed along separate lines: the former focuses on multi-signal information sharing, whereas the latter focuses on residual systematic error learned from negative controls.

In this paper, we develop a Bayesian sequential surveillance framework that integrates empirical bias learning with low-rank latent factor modeling for multi-signal safety surveillance. For each exposure-outcome pair, we decompose the total log-rate-ratio effect into a latent outcome-of-interest effect and an exposure-specific residual bias component. Outcomes used as negative controls inform the exposure-specific bias distribution, while outcomes of interest are modeled jointly through a low-rank latent factor structure that shares information across multiple exposures and outcomes. At each cumulative analysis time, posterior inference is updated using the data accrued to date, producing sequential posterior summaries for latent effects, total effects, and posterior probabilities of increased adverse-event risk. Through simulation studies and a real-data application, we evaluate whether the proposed framework can recover exposure-specific residual bias distributions, provide coherent sequential inference for monitored outcomes, and improve posterior concentration relative to an independent pairwise analysis.

\section{Methods}

This section presents the proposed Bayesian sequential surveillance framework. We first formulate the decomposition of each total exposure-outcome log-rate-ratio into a latent outcome-of-interest effect and an exposure-specific residual bias component. We then describe how outcomes used as negative controls are used for empirical bias learning, and how low-rank latent factors are incorporated to model dependence across multiple exposures and outcomes of interest. Finally, we describe the posterior computation and sequential updating procedure used to propagate learned bias distributions into inference for monitored outcomes.

\subsection{Model formulation}

Suppose there are \(J\) exposures and \(O=m+n\) outcomes. The first \(m\) outcomes are used as negative controls for empirical bias learning, and the remaining \(n\) outcomes are outcomes of interest. Let \(\beta_{j,o}\) denote the total log-rate-ratio effect for exposure \(j=1,\ldots,J\) and outcome \(o=1,\ldots,O\). We decompose this total effect as
\[
\beta_{j,o} = \theta_{j,o} + b_{j,o},
\]
where \(\theta_{j,o}\) is the latent outcome-of-interest effect and \(b_{j,o}\) is the residual systematic bias component.

For outcomes used as negative controls for empirical bias learning, the latent effect is assumed to be zero:
\[
\theta_{j,o}=0,\qquad o=1,\ldots,m.
\]
Therefore, for these outcomes,
\[
\beta_{j,o}=b_{j,o},\qquad o=1,\ldots,m,
\]
so the likelihood contributions from negative controls provide information about the exposure-specific residual bias distribution. For outcomes of interest,
\(o=m+1,\ldots,O\), both the latent effect and the residual bias component are unknown, and inference is based on the decomposition
\[
\beta_{j,o}=\theta_{j,o}+b_{j,o}.
\]

Let
\[
B^{\mathrm{bias}}=[b_{j,o}]_{j=1,\ldots,J;\,o=1,\ldots,O}
\]
denote the \(J\times O\) residual bias matrix. For each exposure \(j\), we assume that the residual bias terms across outcomes arise from an exposure-specific normal distribution,
\[
b_{j,o}\mid \mu_j,\tau_j^2 \sim N(\mu_j,\tau_j^2),
\qquad o=1,\ldots,O,
\]
where \(\mu_j\) and \(\tau_j^2\) characterize the location and variability of the residual systematic bias for exposure \(j\). We assign general hierarchical priors
\[
\mu_j \sim N(0,\sigma_\mu^2),
\qquad
\tau_j^2 \sim \mathrm{Inv\text{-}Gamma}(a,b),
\]
where the hyperparameters are specified in the corresponding analysis setting.

For the outcomes of interest, we introduce a low-rank latent factor structure to share information across exposures and outcomes. Let
\[
\Theta_{\mathrm{OOI}}
=
[\theta_{j,o}]_{j=1,\ldots,J;\,o=m+1,\ldots,O}
\in \mathbb{R}^{J\times n}
\]
denote the latent effect matrix for the outcomes of interest. We model this matrix as
\[
\Theta_{\mathrm{OOI}} = L^{(D)}L^{(O)},
\]
where \(L^{(D)}\in\mathbb{R}^{J\times F}\) is the exposure latent factor matrix, \(L^{(O)}\in\mathbb{R}^{F\times n}\) is the outcome latent factor matrix, and \(F\) is the number of latent factors. 
Thus, for
\(o=m+1,\ldots,O\),
\[
\theta_{j,o}
=
\sum_{f=1}^{F}
L^{(D)}_{j,f}L^{(O)}_{f,o-m}.
\]
We use independent Gaussian priors for the latent factor entries,
\[
L^{(D)}_{j,f}\sim N(0,\sigma_D^2),
\qquad
L^{(O)}_{f,k}\sim N(0,\sigma_O^2),
\]
where \(k=1,\ldots,n\). The full latent effect matrix can then be written as
\[
\Theta =
\left[
0_{J\times m}\mid L^{(D)}L^{(O)}
\right],
\]
and the corresponding total log-rate-ratio matrix is
\[
B=\Theta+B^{\mathrm{bias}}.
\]

This formulation specifies separate prior components for the residual bias matrix and the latent effect structure:
\[
p(B^{\mathrm{bias}},L^{(D)},L^{(O)},\mu,\tau)
=
p(B^{\mathrm{bias}}\mid \mu,\tau)
p(\mu,\tau)
p(L^{(D)})
p(L^{(O)}).
\]
The prior on \(\Theta_{\mathrm{OOI}}\) is therefore induced through the latent factor representation rather than by assigning independent priors directly to each \(\theta_{j,o}\). This factorization separates the empirical bias-learning component from the latent effect component and leads to a convenient posterior computation scheme in which exposure-specific bias distributions are learned from negative controls and then propagated to outcomes of interest.

\subsection{Posterior inference and computation}

Let \(\mathcal{D}_t\) denote the data or likelihood information accrued up to analysis time \(t\). For exposure \(j\) and outcome \(o\), let
\[
L_{j,o,t}(\beta_{j,o})
\]
denote the likelihood contribution for the total log-rate-ratio parameter \(\beta_{j,o}\) at time \(t\). This notation accommodates both individual-level likelihoods and profile-likelihood contributions. In simulation settings, \(L_{j,o,t}(\cdot)\) may be constructed from the individual-level outcome model. In real-data applications, \(L_{j,o,t}(\cdot)\) may be constructed from a profile likelihood evaluated over a grid of log-rate-ratio values, allowing inference to proceed using likelihood summaries without requiring access to individual-level records.

\begin{figure}[h]
\centering
\includegraphics[width=0.9\textwidth]{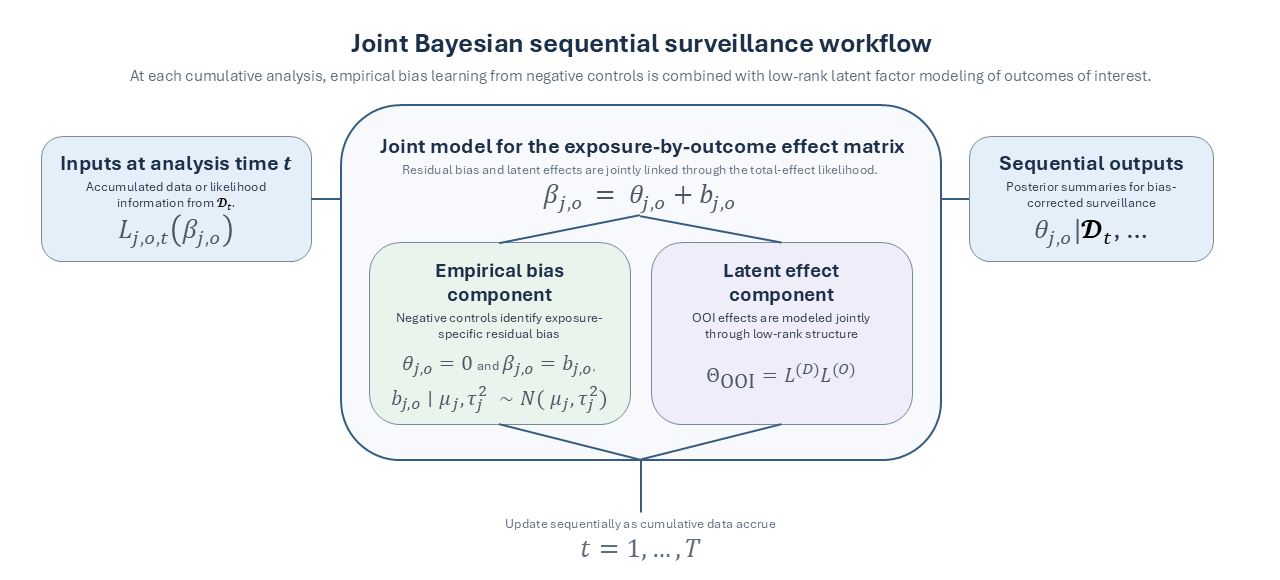}
\caption{Overview of the proposed Bayesian sequential surveillance workflow. At each cumulative analysis time, likelihood information from all monitored exposure-outcome pairs links the empirical bias component and latent outcome-of-interest effect component. Outcomes used as negative controls inform exposure-specific residual bias distributions, while low-rank latent factors model outcomes of interest and share information across correlated signals. The framework produces sequential posterior summaries of \(\theta_{j,o}\) at time \(t\) for bias-corrected surveillance.}
\label{fig:method_flowchart}
\end{figure}

Figure~\ref{fig:method_flowchart} summarizes the proposed Bayesian sequential surveillance workflow at a generic interim analysis time \(t\).

At analysis time \(t\), the likelihood contribution across exposures and outcomes can be written as
\[
L_t(B)
=
\prod_{j=1}^{J}
\left[
\prod_{o=1}^{m}
L_{j,o,t}(b_{j,o})
\prod_{o=m+1}^{O}
L_{j,o,t}(\theta_{j,o}+b_{j,o})
\right].
\]
Combining this likelihood with the separable prior specification gives the posterior distribution
\[
p(B^{\mathrm{bias}},L^{(D)},L^{(O)},\mu,\tau \mid \mathcal{D}_t)
\propto
L_t(B)\,
p(B^{\mathrm{bias}}\mid \mu,\tau)
p(\mu,\tau)
p(L^{(D)})
p(L^{(O)}).
\]
This expression highlights the two sources of information in the model: negative controls inform the exposure-specific bias distributions, whereas the outcomes of interest inform the latent factor structure after accounting for residual bias.

For computation, we exploit this factorization. First, the negative control outcomes are used to learn the exposure-specific bias distributions. Because \(\theta_{j,o}=0\) for \(o=1,\ldots,m\), the likelihood for these outcomes depends on \(b_{j,o}\) through \(\beta_{j,o}=b_{j,o}\). The posterior for the bias-learning component is
\[
p(B^{\mathrm{bias}}_{\mathrm{NC}},\mu,\tau \mid \mathcal{D}_t)
\propto
\left[
\prod_{j=1}^{J}\prod_{o=1}^{m}
L_{j,o,t}(b_{j,o})
\right]
\left[
\prod_{j=1}^{J}\prod_{o=1}^{m}
p(b_{j,o}\mid \mu_j,\tau_j^2)
\right]
p(\mu,\tau),
\]
where \(B^{\mathrm{bias}}_{\mathrm{NC}}\) denotes the residual bias terms for
the negative control outcomes. Posterior samples of
\[
\left\{\mu_j^{(s)},\tau_j^{2(s)}\right\}_{s=1}^{S},
\qquad j=1,\ldots,J,
\]
summarize the empirical bias distribution for each exposure. These posterior draws can also be saved as empirical bias distributions for the same exposure set and data source and propagated to additional outcomes of interest.

Second, posterior draws from the learned exposure-specific bias distributions are propagated into inference for the outcomes of interest. For posterior draw \(s\), bias terms for the outcomes of interest are sampled as
\[
b_{j,o}^{(s)}
\sim
N\left(\mu_j^{(s)},\tau_j^{2(s)}\right),
\qquad o=m+1,\ldots,O.
\]
Conditional on these bias draws, the latent factor matrices \(L^{(D)}\) and \(L^{(O)}\) are updated using the likelihood contributions from the outcomes of interest:
\[
\prod_{j=1}^{J}\prod_{o=m+1}^{O}
L_{j,o,t}
\left(
\sum_{f=1}^{F}
L^{(D)}_{j,f}L^{(O)}_{f,o-m}
+
b_{j,o}^{(s)}
\right).
\]
Posterior computation uses Markov chain Monte Carlo updates. The bias-learning component can be sampled using Gibbs and Metropolis-Hastings updates, and the latent factor component is sampled using Metropolis-Hastings updates for
\(L^{(D)}\) and \(L^{(O)}\).

At each analysis time \(t\), inference is performed using the likelihood constructed from all information accrued up to \(\mathcal{D}_t\). For each outcome of interest, posterior draws of the total log-rate-ratio effect are obtained as
\[
\beta_{j,o}^{(s)}
=
\theta_{j,o}^{(s)}+b_{j,o}^{(s)},
\qquad o=m+1,\ldots,O.
\]
Posterior summaries are then computed from the corresponding posterior draws.

\section{Simulation Study}

We conducted simulation studies to evaluate whether the proposed framework can recover exposure-specific residual bias distributions and provide coherent sequential inference for latent outcome-of-interest effects. The simulation design was constructed to resemble key features of postmarket safety surveillance, including longitudinal exposure histories, multiple monitored exposures and outcomes, outcomes used as negative controls for empirical bias learning, and repeated cumulative analyses as information accrues. Because the data-generating parameters are known in simulation, we used this setting to assess both the sequential posterior behavior of latent effects for outcomes of interest and the bias recovery.

\subsection{Simulation design}

We generated longitudinal data for \(N=600\) patients over a one-year observation period. The simulation included \(J=4\) exposures and \(O=55\) outcomes, with \(m=50\) outcomes used as negative controls for empirical bias learning and \(n=5\) outcomes of interest. The latent factor dimension was set to \(F=2\). A complete dataset was generated first, and cumulative analyses were performed at 12 equally spaced monthly analysis times.

\begin{table}[htbp]
\centering
\caption{Simulation design settings.}
\label{tab:sim_design}
\begin{tabular}{lc}
\hline
Setting & Value \\
\hline
Number of patients & \(N=600\) \\
Observation window & One year \\
Number of cumulative analyses & 12 monthly analyses \\
Number of exposures & \(J=4\) \\
Number of outcomes & \(O=55\) \\
Outcomes used for empirical bias learning & \(m=50\) \\
Outcomes of interest & \(n=5\) \\
Latent factor dimension & \(F=2\) \\
\hline
\end{tabular}
\end{table}

For each exposure \(j\), we generated exposure-specific bias parameters
\[
\mu_j \sim N(0,0.2^2), \qquad \tau_j \sim \mathrm{Uniform}(0.1,0.3),
\]
and then generated
\[
b_{j,o} \sim N(\mu_j,\tau_j^2)
\]
for all outcomes. For the outcomes used as negative controls, \(\theta_{j,o}=0\). For the outcomes of interest, the latent effect matrix was generated from a rank-\(F\) structure,
\[
\Theta_{\mathrm{OOI}} = L^{(D)}L^{(O)}.
\]
The total log-rate-ratio matrix was then defined as
\[
B = \Theta + B^{\mathrm{bias}}.
\]

Patient-level exposure histories were generated over the observation period. Each patient was assigned a random subset of exposures, and each selected exposure was assigned a finite exposure window. Exposure window lengths were sampled from 3 to 7 time points. Partial overlaps between exposure windows were allowed, while complete containment of one exposure window within another was avoided. Conditional on the simulated exposure history and total log-rate-ratio matrix, outcome counts were generated from
\[
Y^{(o)}_{i,d} \sim \mathrm{Poisson}\left\{\exp\left(x_{i,d}^{\top}\beta^{(o)}\right)\right\},
\]
where \(x_{i,d}\) is the exposure vector for patient \(i\) at time point \(d\), and \(\beta^{(o)}=(\beta_{1,o},\ldots,\beta_{J,o})^\top\).

\subsection{Sequential posterior behavior of latent outcome-of-interest effects}

We first examined posterior behavior for latent outcome-of-interest effects across monthly analyses. For each exposure-outcome-of-interest pair, posterior draws of \(\theta_{j,o}\) were summarized using half-violin plots, posterior medians, and 95\% credible intervals. These plots evaluate how posterior location and uncertainty evolve as information accumulates.

\begin{figure}[htb]
\centering
\includegraphics[width=0.85\textwidth]{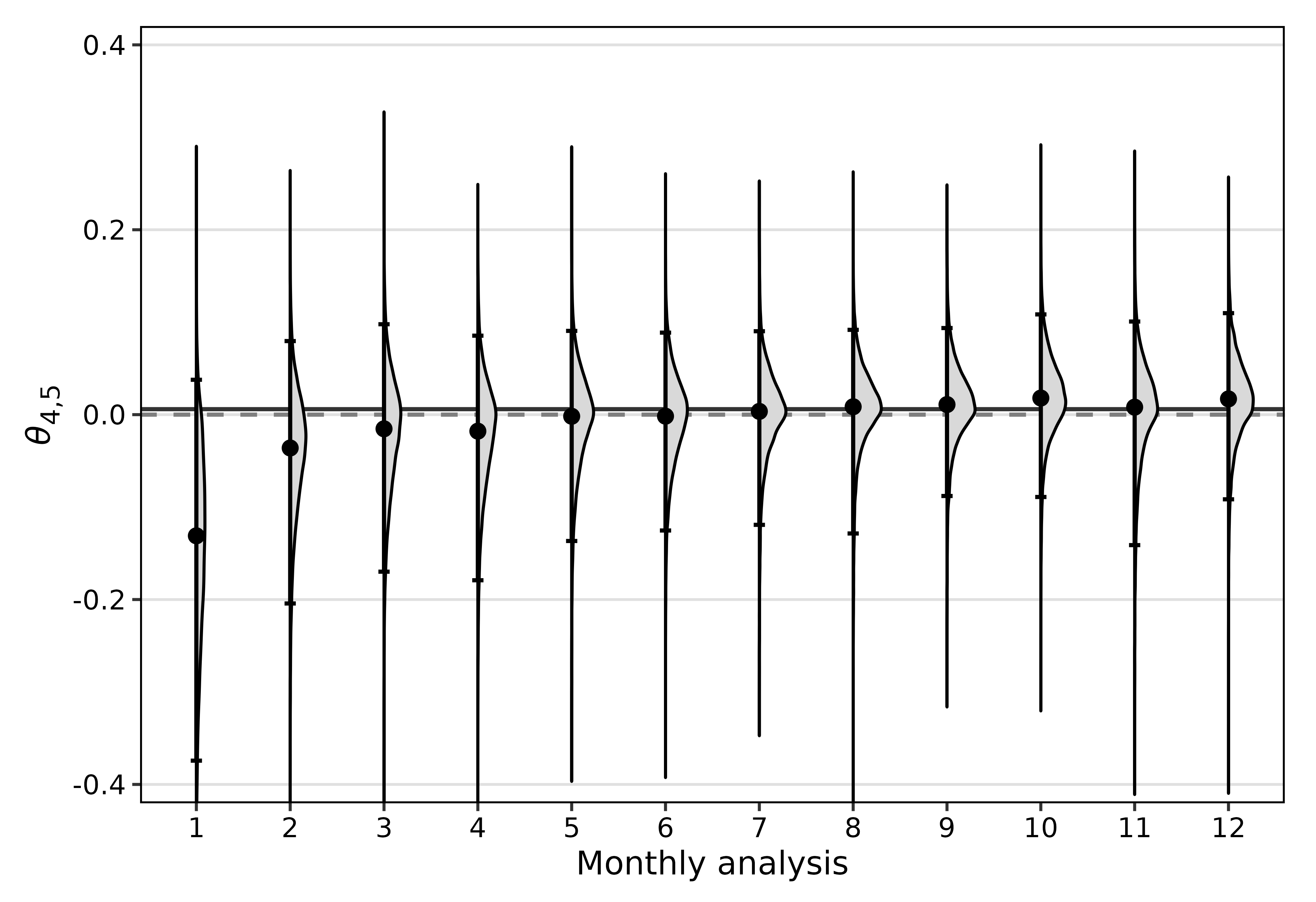}
\caption{Sequential posterior summaries for a representative latent outcome-of-interest effect in the simulation study. Half-violins show posterior distributions of \(\theta_{4,5}\) across monthly analyses. Points indicate posterior medians, vertical intervals indicate 95\% credible intervals, the solid horizontal line marks the true value of \(\theta_{4,5}\), and the dashed horizontal line marks the null effect \(\theta=0\).}
\label{fig:sim_theta}
\end{figure}

Figure~\ref{fig:sim_theta} illustrates the sequential posterior behavior for a representative latent outcome-of-interest effect. In the earliest monthly analyses, the posterior distribution is relatively diffuse, reflecting limited accumulated information. As additional data accrue, posterior uncertainty decreases and the posterior summaries become more stable across analyses. Although uncertainty remains substantial, the posterior medians move closer to the true value after the earliest analyses. This pattern illustrates how the proposed model sequentially updates inference for latent outcome-of-interest effects as evidence accumulates over time.

\subsection{Recovery of exposure-specific bias distributions}

We next evaluated whether the bias-learning procedure could recover the exposure-specific bias distributions from the outcomes used as negative controls for empirical bias learning. For each exposure, we compared the posterior-averaged learned bias density with the corresponding true bias density. The true density was parameterized by the true values of \(\mu_j\) and \(\tau_j\), and the learned density at each monthly analysis was obtained by averaging normal densities over posterior samples of \(\mu_j\) and \(\tau_j\). Thus, the learned curves represent posterior-averaged estimates of the underlying bias distribution rather than plug-in densities based on a single posterior summary.

\begin{figure}[htb]
\centering
\includegraphics[width=\textwidth]{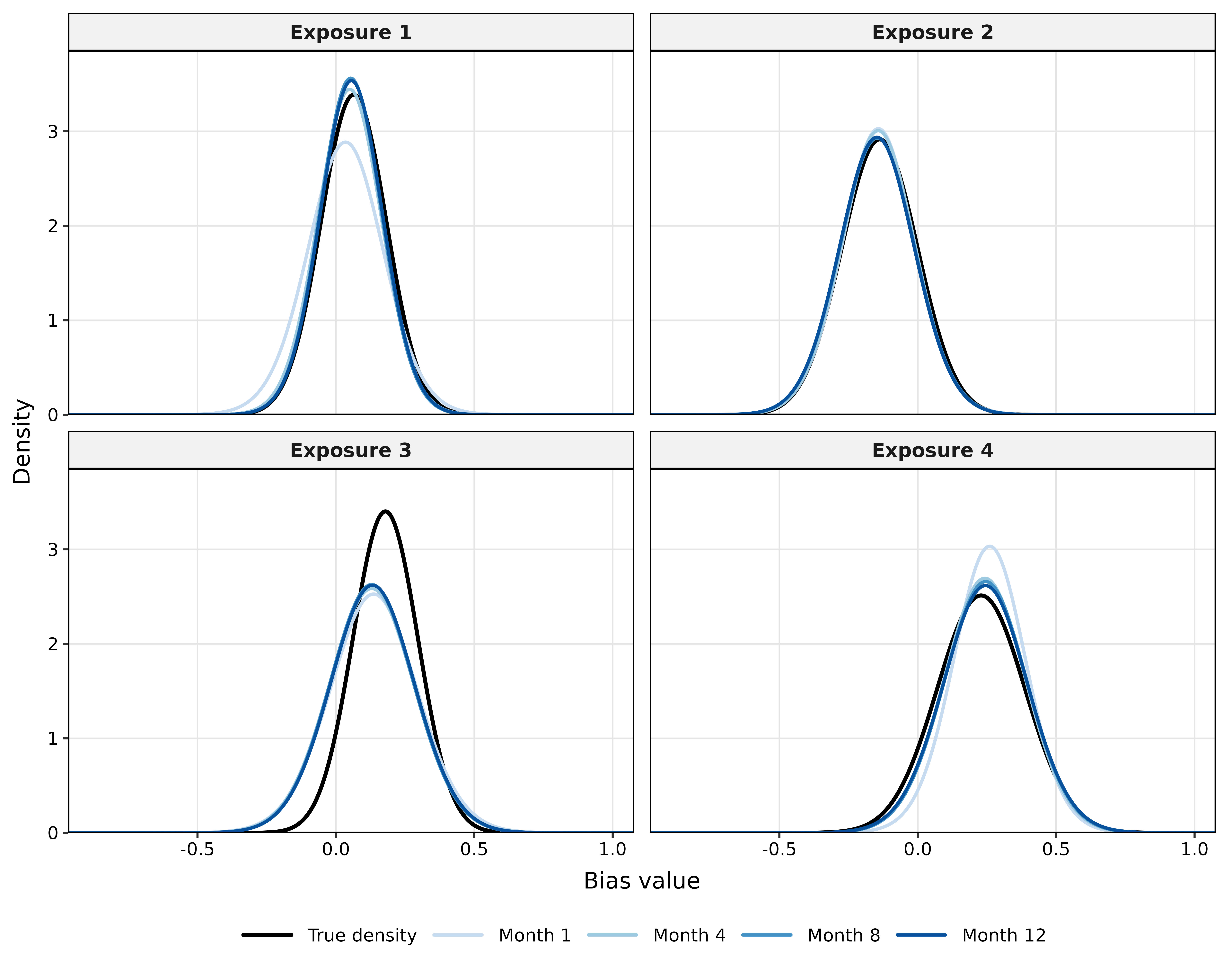}
\caption{Recovery of exposure-specific bias distributions in the simulation study. For each exposure, the black curve represents the true bias density determined by \(\mu_j\) and \(\tau_j\). Colored curves show posterior-averaged learned bias densities at Months 1, 4, 8, and 12.}
\label{fig:sim_bias}
\end{figure}

Figure~\ref{fig:sim_bias} shows that the learned bias distributions generally moved toward the corresponding true bias distributions as additional monthly analyses were incorporated. For Exposures 1, 2, and 4, the posterior-averaged learned densities became increasingly aligned with the true densities over time, indicating improved recovery of exposure-specific bias distributions as evidence accumulated. For Exposure 3, although recovery was less complete than for the other exposures, the learned density still showed a clear shift toward the corresponding true density across monthly analyses. Overall, these results indicate that the bias-learning procedure can recover exposure-specific residual bias distributions with reasonable accuracy as sequential information accumulates.

\section{Real-Data Application}

The real-data application is intended to demonstrate the practical use of the proposed framework in a setting where true exposure-outcome effects are unknown. We focus on how empirical bias correction and latent factor modeling affect posterior concentration, sequential updating, and posterior directional evidence across multiple monitored signals. Accordingly, the analysis is presented as a methodological illustration of bias-corrected multi-signal surveillance rather than as a signal-discovery exercise.

\subsection{Data source and analysis setting}

We applied the proposed framework to a postmarket vaccine safety surveillance analysis using profile likelihoods from a large US insurance claims database \citep{database}. The analysis used a historical comparator design and was conducted over a series of cumulative analyses. At each analysis period, the profile likelihoods represented accumulated evidence up to that period, allowing posterior summaries to be updated as additional data accrued.

The real-data analysis used profile likelihood inputs rather than individual-level patient records. For each exposure-outcome pair and cumulative analysis period, an interpolated profile log-likelihood was constructed as a function of the total log-rate-ratio parameter. These likelihoods were then incorporated into the Bayesian bias-correction and latent factor modeling framework described above.

\subsection{Selection of exposures and outcomes}

To ensure comparability across cumulative analyses, we fixed the exposure and outcome sets before fitting the model. Exposures were retained if profile likelihoods were available across all analysis periods. Outcomes were retained if profile likelihoods were available for all fixed exposures across the same period range.

The real-data application included four vaccine exposure groups: H1N1pdm, seasonal flu vaccine Fluvirin, HPV vaccine Gardasil 9, and zoster vaccine Shingrix. Five validation outcomes were selected as outcomes of interest for the real-data illustration. These outcomes are labeled as OOI 1--OOI 5 in the figures, with detailed outcome descriptions and standardized vocabulary concept identifiers provided in the Supplementary Material.

The remaining fixed outcomes were used as negative controls for empirical bias learning. Thus, the same monitored set was used across all cumulative analyses, allowing changes in posterior summaries over time to reflect sequentially accumulated evidence rather than changes in the monitored exposure or outcome set.

\subsection{Posterior concentration under latent factor modeling}

We compared the proposed latent factor model with a pairwise baseline. Both analyses used the same fixed exposure-outcome set and the same empirical bias-learning procedure. The primary latent factor analysis used rank \(F=2\). The only difference was in the model for outcomes of interest: the proposed method imposed a low-rank latent factor structure on the outcome-of-interest effect matrix, whereas the baseline estimated each exposure-outcome effect independently. 

\begin{figure}[p]
\centering
\includegraphics[width=\textwidth,height=0.9\textheight,keepaspectratio]{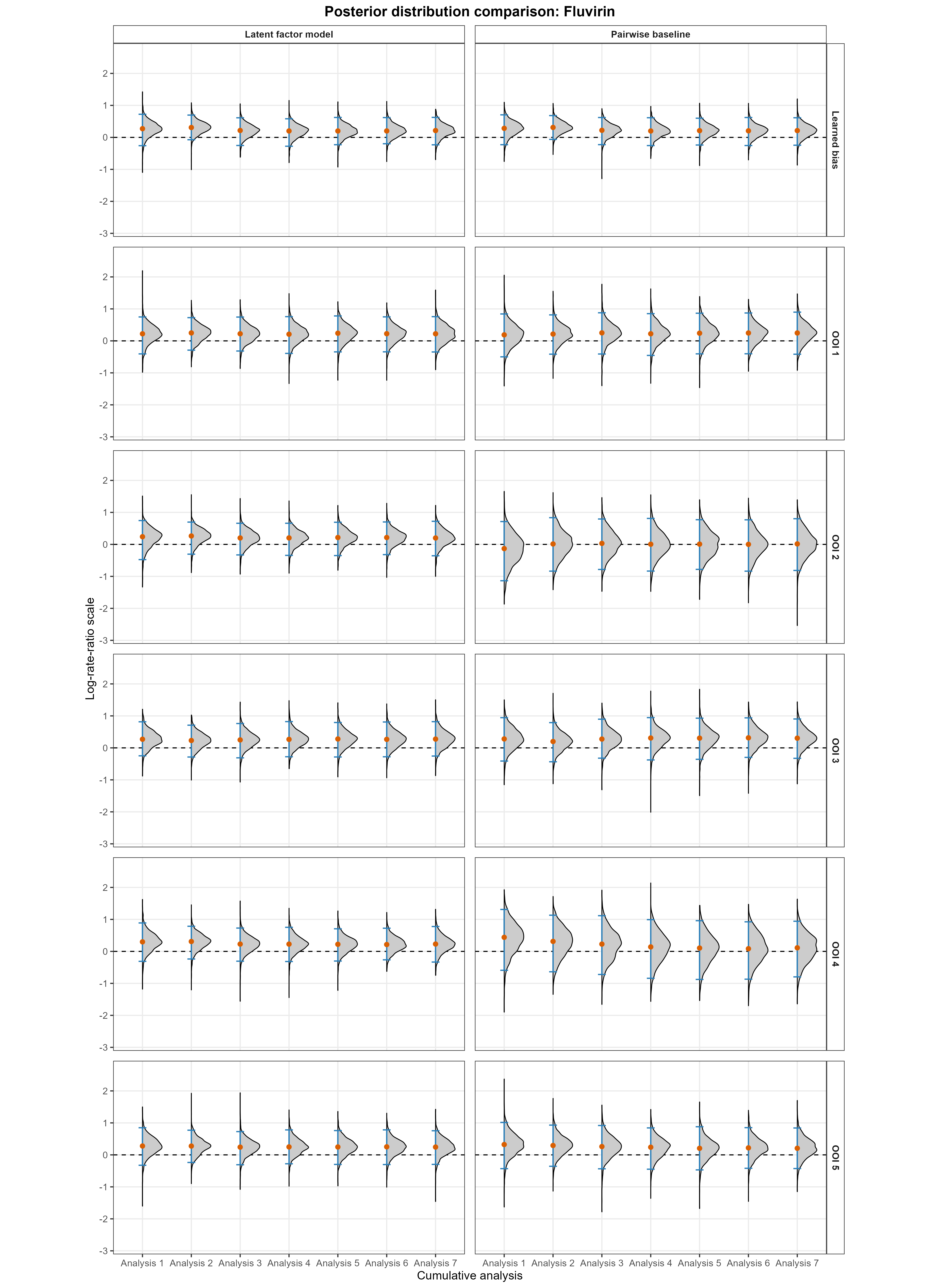}
\caption{Posterior distribution comparison with and without latent factor modeling for the seasonal flu vaccine Fluvirin exposure group. Both analyses used the same fixed exposure-outcome set and the same empirical bias-learning procedure. Points indicate posterior means, intervals indicate 95\% credible intervals, and the dashed horizontal line marks zero.}
\label{fig:real_compare_fluvirin}
\end{figure}

Figure~\ref{fig:real_compare_fluvirin} illustrates this comparison for the seasonal flu vaccine Fluvirin exposure group. The left column shows results from the proposed latent factor model, while the right column shows results from the pairwise baseline. Rows correspond to the learned bias distribution and the five outcomes of interest, and columns within each panel correspond to cumulative analyses.

Across the four exposure groups, the learned bias distributions were broadly similar under the latent factor and pairwise models, as expected because the empirical bias-learning procedure was shared. The main differences therefore arise from the model for outcomes of interest. Relative to the pairwise baseline, the latent factor model generally produced posterior distributions that were more concentrated and more stable across cumulative analyses, with less fluctuation in posterior means and interval widths over time.

For the Fluvirin exposure group, the learned bias row is nearly unchanged between the two analyses, indicating that both approaches begin from essentially the same empirical bias-learning result. In contrast, the outcome-of-interest posterior distributions are systematically tighter under the latent factor model. This pattern is visible across multiple validation outcomes, where the pairwise baseline shows wider or less regular posterior distributions across cumulative analyses. These results suggest that, for the Fluvirin exposure group, structured borrowing through the latent factor model improves the efficiency of inference for the outcomes of interest.

\subsection{Endpoint posterior precision comparison}

We further summarized posterior uncertainty at the final cumulative analysis. For each exposure group, we computed the 95\% credible interval width for each of the five outcomes of interest and reported the mean and standard deviation across outcomes.

\begin{table}[htbp]
\centering
\caption{Endpoint posterior precision comparison. For each exposure group, values are mean (SD) of the 95\% credible interval widths across the five outcomes of interest at the final cumulative analysis. Relative reduction is computed as the percentage decrease in mean interval width under the latent factor model relative to the pairwise baseline.}
\label{tab:endpoint_precision}
\begin{tabular}{lccc}
\hline
Exposure group & Latent factor model & Pairwise baseline & Relative reduction \\
\hline
H1N1pdm      & 1.899 (0.078) & 2.217 (0.017) & 14.4\% \\
Fluvirin     & 1.084 (0.024) & 1.434 (0.230) & 24.4\% \\
Gardasil 9 & 4.796 (0.147) & 4.695 (0.113) & -2.2\% \\
Shingrix   & 2.450 (0.158) & 2.625 (0.166) & 6.7\% \\
\hline
\end{tabular}
\end{table}

At the final cumulative analysis, the latent factor model yielded narrower average credible intervals for three of the four exposure groups. The largest reduction was observed for the seasonal flu vaccine Fluvirin group, where the mean interval width decreased from 1.434 under the pairwise baseline to 1.084 under the latent factor model, corresponding to a 24.4\% reduction. Smaller reductions were observed for H1N1pdm and zoster vaccine Shingrix. For HPV vaccine Gardasil 9, the two approaches produced similar average interval widths, indicating that posterior precision remained stable even though this exposure group contributed little to the overall efficiency gain. Overall, these results suggest that the latent factor component can substantially improve posterior concentration when cross-outcome borrowing is informative, while preserving stable uncertainty when such borrowing is less pronounced.

\subsection{Posterior probability summaries over cumulative analyses}

We examined posterior probabilities of positive total effects over cumulative analyses as a surveillance summary. For each exposure group and outcome of interest, we computed
\[
\Pr(\beta_{j,o} > 0 \mid D_t),
\]
where \(\beta_{j,o}\) denotes the total log-rate-ratio effect and \(D_t\) denotes the data accumulated up to cumulative analysis \(t\). These posterior probabilities summarize the direction and strength of evidence on the total-effect scale after empirical bias correction and latent factor modeling. From a Bayesian testing perspective, \(\Pr(\beta_{j,o}>0\mid D_t)\) can be viewed as the posterior probability of the increased-risk alternative \(H_1:\beta_{j,o}>0\), relative to the no-increased-risk hypothesis \(H_0:\beta_{j,o}\le 0\). Thus, in safety surveillance, larger posterior probabilities provide stronger evidence that an exposure may be related to increased adverse-event risk.

\begin{figure}[htbp]
\centering
\includegraphics[width=\textwidth]{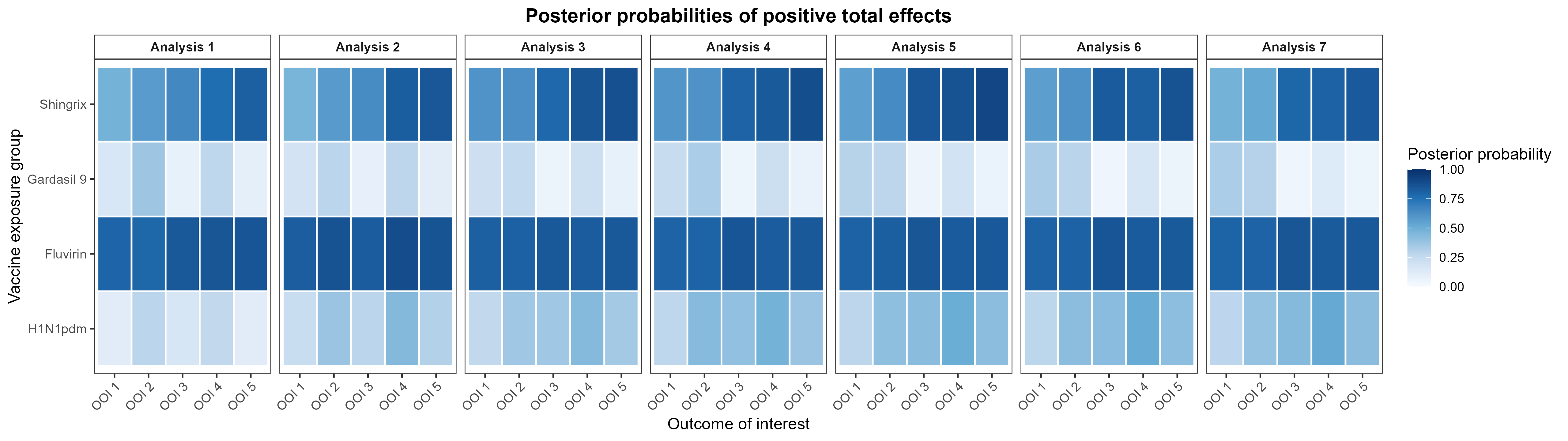}
\caption{Posterior probabilities of positive total effects across cumulative analyses in the real-data application. Each cell shows \(\Pr(\beta_{j,o}>0 \mid \mathcal{D}_t)\) for an exposure group and outcome of interest, where \(\beta_{j,o}\) denotes the total log-rate-ratio effect. Outcomes are labeled as OOI 1--OOI 5, with detailed descriptions provided in the Supplementary Material.}
\label{fig:real_heatmap}
\end{figure}

Figure~\ref{fig:real_heatmap} summarizes how posterior directional evidence evolved as additional data accrued. For most monitored exposure-outcome pairs, posterior probabilities remained close to 0.5 or showed moderate departures from 0.5, consistent with the intended role of these outcomes in the real-data analysis. This pattern supports their use for evaluating posterior behavior and sequential updating, rather than for highlighting specific exposure-outcome associations.

The heatmap also provides a compact view of sequential updating across cumulative analyses. For exposure-outcome pairs with posterior probabilities that remained away from 0.5, the direction of evidence was generally maintained across analyses. Together with the posterior distribution comparison and endpoint precision summary, these results show that the proposed framework can provide interpretable sequential summaries of both posterior uncertainty and directional evidence in a real-world surveillance setting.

We repeated the real-data analysis using \(F=3\) and \(F=4\), with \(F=4\) corresponding to the maximum possible rank for the \(4\times 5\) exposure-by-outcome-of-interest matrix. Posterior summaries under \(F=3\) were close to those from the primary \(F=2\) analysis, with similar posterior mean estimates, credible interval widths, and posterior probabilities of positive total effects. The rank-\(4\) analysis retained the main qualitative posterior probability patterns while allowing greater flexibility in the latent effect structure. Across these sensitivity analyses, the primary conclusions were stable, and the rank-\(2\) latent factor model provided a parsimonious low-rank representation that preserved posterior concentration and interpretability. Detailed rank-sensitivity summaries are provided in the Supplementary Material.

\section{Discussion}

We proposed a multi-signal Bayesian sequential surveillance framework that combines empirical bias correction from outcomes used as negative controls with low-rank latent factor modeling of the exposure-by-outcome effect matrix. The central contribution is to bring together two ideas: empirical learning of exposure-specific residual bias distributions and structured borrowing of information across multiple exposures and multiple outcomes  \citep{schuemie2014interpreting, schuemie2016robust, tchetgen2014control, bu2024bayesian, moghaddass2016factorized}. This allows posterior inference to account for systematic error while also improving precision through shared latent structure. The framework is particularly suited to postmarket safety surveillance settings, where many exposure-outcome pairs must be monitored repeatedly as evidence accumulates over time \citep{lieu2007real, kulldorff2011maximized, li2020bayesian, bu2024bayesian}.

The empirical evaluations illustrate the practical utility of this framework. In the simulation study, the model was designed to assess whether exposure-specific bias distributions can be learned from negative controls and whether posterior inference for outcomes of interest behaves coherently over sequential analyses. In the real-data application, the same framework was applied to a complex vaccine safety surveillance setting using profile likelihood summaries from a large observational healthcare database \citep{database}. Together, these analyses demonstrate how the proposed model can combine empirical bias learning, sequential updating, and multi-signal information sharing within a unified inferential workflow. Importantly, the real-data application should not be interpreted primarily as a positive signal-discovery analysis. The selected outcomes were used as validation outcomes for which strong positive effects were
not expected. Thus, posterior effects that remain close to the null provide empirical support for the robustness of the bias-correction procedure, while the latent factor structure can improve posterior precision through borrowing across related exposure-outcome pairs \citep{li2020bayesian, bu2024bayesian}.

Several limitations suggest directions for future work. The bias-learning component relies on an exchangeability assumption for the outcomes used as negative controls within each exposure, which may be restrictive if residual systematic error differs substantially across negative controls \citep{schuemie2014interpreting, schuemie2016robust, tchetgen2014control}. In addition, the exposure-specific normal bias distribution may not fully accommodate skewness, heavy tails, or multimodality; more flexible mixture distributions provide a natural extension. The latent factor rank \(F\) is fixed in the current implementation. Although our rank-sensitivity analyses suggested that the main conclusions were stable, formal rank selection or posterior uncertainty quantification over the latent dimension remains an important methodological extension \citep{ghosh2009default, lopes2004bayesian, bhattacharya2011sparse, legramanti2020bayesian}. Other extensions may include more scalable posterior computation, synthesis across multiple databases, and time-varying latent factor structures for larger surveillance networks \citep{madigan2014systematic, schuemie2018improving}.

\section*{Conflicts of Interest}

The authors declare no conflicts of interest.

\section*{Data Availability}

The data that support the findings of this study are available on request from the corresponding author. The data are not
publicly available due to privacy or ethical restrictions.

\section*{Code Availability}

Code implementing the proposed methodology, together with the simulation and analysis scripts used to reproduce the results in this article, is publicly available at GitHub: \url{https://github.com/ZiyangPan/BayesianSurveillance}.

\printendnotes

\bibliographystyle{WileyNJD-AMA}
\bibliography{references}

\end{document}